# Specification description and verification of multitask hybrid systems in the OTS/CafeOBJ method

Masaki Nakamura and Kazutoshi Sakakibara

Department of Information Systems Engineering, Faculty of Engineering, Toyama Prefectural University, Toyama, Japan

Kazuhiro Ogata

Japan Advanced Institute of Science and Technology, Ishikawa, Japan

## Abstract

To develop IoT and/or CSP systems, we need consider both continuous data from physical world and discrete data in computer systems. Such a system is called a hybrid system. Because of density of continuous data, it is not easy to do software testing to ensure reliability of hybrid systems. Moreover, the size of the state space increases exponentially for multitask systems. Formal descriptions of hybrid systems may help us to verify desired properties of a given system formally with computer supports. In this paper, we propose a way to describe a formal specification of a given multitask hybrid system as an observational transition system in CafeOBJ algebraic specification language and verify it by the proof score method based on equational reasoning implemented in CafeOBJ interpreter.

# 1 Introduction

Formal methods are mathematically based techniques for specification and verification of software and hardware systems. Formal specification languages play important role in formal methods. CafeOBJ is an executable formal specification language, which provides specification execution based on a rewrite theory [1]. The OTS/CafeOBJ method is a formal method in which a system is modeled as an observational transition system (OTS), its specification is described in CafeOBJ, and properties are verified formally by using specification execution function implemented in CafeOBJ, called the proof score method [2, 3, 4]. In our previous work [5, 6], we have proposed a way to describe and verify multitasking real-time systems in the OTS/CafeOBJ method. A real-time system is regarded as a hybrid system where only time is its continuous data. In the literature [7], a way to describe and verify formal specifications of hybrid systems based on CafeOBJ has been proposed, however, only a single-task system is considered. In this study, we propose a way to describe a formal specification of a multitask hybrid system as an observational transition system in CafeOBJ algebraic specification language by extending the existing results, and verify it by the proof score method based on equational reasoning implemented in CafeOBJ interpreter (Table 1).

Table 1: A position of this study

|  | real-time system | hybrid system |
|---|---|---|
| single task | [4] | [7] |
| multitask | [5, 6] | This study |

The remainder of this paper is structured as follows. In Section 2, we introduce the OTS/CafeOBJ method, including description of data and system specifications and verification with rewriting. We propose a way to describe specifications for multitask hybrid systems with an arbitrary number of processes in Section 3. In Sections 4 and 5, we simulate and verify described specifications by equational reasoning in CafeOBJ interpreter. We discuss about related work in Section 6 and conclude our work in Section 7.

# 2 Preliminaries

We introduce some notion and notation of CafeOBJ specification language and system [1] and the OTS/CafeOBJ method including the proof score method [2, 3].

## 2.1 CafeOBJ modules

A CafeOBJ specification consists of modules. A CafeOBJ module $\mathsf{SP} = (\Sigma_{\mathsf{SP}}, E_{\mathsf{SP}})$ consists of its signature and axioms. A signature $\Sigma_{\mathsf{SP}} = (S_{\mathsf{SP}}, \leq_{\mathsf{SP}}, F_{\mathsf{SP}})$ consists of a set of sorts, an ordering on the sorts and an $S_{\mathsf{SP}}$-sorted set of operations. A $\Sigma_{\mathsf{SP}}$-algebra $A$ is an algebra which has a carrier set $A_s$ for each $s \in S_{\mathsf{SP}}$ and an operation $A_f : A_{s_0} \times \cdots \times A_{s_n} \to A_s$ for each $f \in (F_{\mathsf{SP}})_{s_0 \ldots s_n s}$.

An axiom $l = r$ if $c \in E$ is a conditional equation whose both sides $l$ and $r$ of the equation are terms of a same sort and condition is a term of boolean sort constructed from the operations in $\Sigma_{SP}$ and variables. A SP-algebra is a $\Sigma_{SP}$-algebra which satisfies all equations in $E_{SP}$. When SP has tight denotation, it denotes the initial SP-algebra. When SP has loose denotation, it denotes all SP-algebras. See [1] for more details.

The following is a loose module specifying an arbitrary set with a binary predicate.

```
mod* PID {
  [ Pid ]
  op _=_ : Pid Pid -> Bool {comm}
}
```

The module declared with `mod*` has loose denotation, which denotes all models satisfying axioms. Module PID has sort Pid and binary operation _=_, and has no axioms, where the underlines indicate the position of arguments in term expression, that is, $t = t'$ is a term of Bool for terms $t, t'$ of Pid. Module PID denotes (the set of) all algebras with a binary predicate, which includes Boolean algebra and algebras of natural numbers, integers, real numbers, etc.

The following is a tight module specifying labels of a traffic signal.

```
mod! LABEL{
  [Label]
  ops green red yellow : -> Label
  pred _=_ : Label Label {comm}
  op next : Label -> Label
  var L : Label
  eq (L = L) = true .
  eq (green = red) = false .
  eq (green = yellow) = false .
  eq (red = yellow) = false .

  eq next(red) = green .
  eq next(green) = yellow .
  eq next(yellow) = red .
}
```

The module declared with `mod!` has tight denotation, which denotes the initial model. In the initial mode, any elements of a carrier set is represented by a term constructed from its signature, and no two elements of a carrier set are equivalent unless the corresponding terms can be shown to be equal using its axioms. Module LABEL has sort Label. Operations green, yellow, red with the empty arity are declared. Such an operation is interpreted as a constant element of a carrier set. Operation _=_ is a binary predicate. The first four equations define the equality predicate, which takes two labels and returns true if they are same, otherwise false. Note that the first equation (L = L) = true specifies that term $t = t$ is equivalent to true for each term $t$ of Label. Operation next is an unary operation. The next label is defined like red $\Rightarrow$ green $\Rightarrow$ yellow $\Rightarrow$ red. Since $t = t'$ and next($t$) are well-defined for all labels $t, t'$, the tight module LABEL denotes the algebra whose carrier set has exactly three elements.

## 2.2 Equational reasoning in CafeOBJ

CafeOBJ supports a reduction command to do equational reasoning. The reduction command regards an equation as a left-to-right rewrite rule. Let $t$ be a term. When $t$ is reduced into $t'$ by using equations $E_{SP}$ in module SP, equation $t = t'$ is deducible from $E_{SP}$, denoted by $E_{SP} \vdash t = t'$, and the proposition $t = t'$ holds for all models of SP.

The followings are examples of reduction:

```
red in LABEL : next(next(green)) .
red in LABEL : next(green) = next(yellow) .
```

Since the first reduction returns `red`, we have $E_{LABEL} \vdash \texttt{next(next(green))} = \texttt{red}$. The second reduction returns `false`. We have $E_{LABEL} \vdash (\texttt{next(green)} = \texttt{next(yellow)}) = \texttt{false}$.

By opening a module, we can add operations and equations to the module. When module SP is open and a fresh constant op *elem* : -> *sort* is declared, we can use *elem* as an arbitrary element of $S_{sort}$. Consider term $t$ and equation $l = r$ constructed from the signature and *elem*. If term $t$ is reduced into $t'$ after declaring $l = r$, $E_{SP} \cup \{l = r\} \vdash t = t'$ holds, that is, implication $l = r \Rightarrow t = t'$ holds for an arbitrary element *elem* in all models of SP. The proof score method consists of those kinds of proof passages to make a proof of a given property by using proof techniques e.g. case splitting and induction. The following is an example of a proof passage:

```
open LABEL .
  op s : -> Label .
  red next(next(next(s))) = s .
close .
```

A declared fresh constant `s` of sort `Label` denotes an arbitrary element in the following reduction. The reduction command means that the next of next of next of an arbitrary label is equivalent with the label itself. Although it is true, CafeOBJ does not prove it directly. The above reduction returns the input term as it is. For this example, case splitting is useful.

```
open LABEL .
  op s : -> Label .
  eq s = green .
  red next(next(next(s))) = s .
close .
open LABEL .
  op s : -> Label .
  eq s = yellow .
  red next(next(next(s))) = s .
close .
open LABEL .
  op s : -> Label .
  eq s = red .
  red next(next(next(s))) = s .
close .
```

The all three reductions return `true` and the case splitting `s = green` ∨ `s = yellow` ∨ `s = red` is complete since `LABEL` is tight. Therefore, the above combination of three proof passages is regarded as a formal proof of $\forall s \in A_{\texttt{Label}}.A_{\texttt{next}}(A_{\texttt{next}}(A_{\texttt{next}}(s))) = s$. Such a construction of complete formal proofs is called the proof score method.

CafeOBJ also supports built-in modules for typical data types. A built-in module consists of a signature and its model, instead of axioms. For example, RAT denotes an algebra of rational numbers. In RAT, term 1/2 + 3/4 is reduced into 5/4.

### 2.3 The OTS/CafeOBJ method

An OTS/CafeOBJ specification consists of data modules and a system module. Modules LOC and PID are examples of data modules. A system module is given as a behavioral specification of CafeOBJ. A behavioral specification has a special sort, called a hidden sort, and special operations, called behavioral operations, whose arguments include the hidden sort. A behavioral operation whose returned sort is not hidden is called an observation. A behavioral operation whose returned sort is hidden is called a transition. Two elements of the hidden sort are observationally equivalent if their observed values are equivalent for each observation. An OTS/CafeOBJ specification is a restricted behavioral specification, where observational equivalence is preserved by transitions.

The following is an OTS/CafeOBJ specification of a signal control system:

```
mod* SIGNAL{
  pr(LABEL)
  *[Sys]*
   op init : -> Sys
  bop color : Sys -> Label
  bop change : Sys -> Sys
  var S : Sys
  eq color(init) = green .
  eq color(change(S)) = next(color(S)) .
}
```

Figure 1 represents the OTS model described in SIGNAL. Module SIGNAL imports module LABEL with the protecting mode, where a model of the importing module includes a model of the imported module as it is. Hidden sort Sys is declared, which denotes the state space of a system to be specified. Constant init is declared as an initial state. Observation color observes a color, where term color($s$ represents the current color in state $s$ of the signal control system. Transition change changes a color, where term change($s$ represents the state obtained after changing.

The first equation color(init) = green specifies the initial state through observation such that the initial color is green. The second equation specifies the behavior of transition change through observation. Term change(S) represents the state obtained by applying change to S. The color of change(S) is defined as the next color of S. For example, term color(change(change(init))) is reduced into red.

## 3 OTS/CafeOBJ specifications of hybrid systems

In this section, we introduce a way to describe OTS/CafeOBJ specifications of hybrid systems. We model signal control systems with a single car and with plural cars as hybrid automata, and describe them as OTS/CafeOBJ specifications.

Figure 1: An OTS model of a signal

## 3.1 A hybrid automaton of a signal control system

A simple signal control system, represented in Fig.2, consists of a signal and a car such that the car is prohibited from being in a specific area, between $cs_0$ and $cs_1$, while the signal is red. The specific area is called the critical section. The signal has three modes indicating its color label. Each color of the signal should be kept more than $t_0$ time units. The car has two modes: going and not-going. In the going mode, the car moves forward according time advancing. The car stays in the not-going mode. If the signal label is not green, the car cannot enter the critical section. If the signal label is changed into yellow while the car exists in the critical section, the car should keep the going mode, that is, it should not stop. Thus, if the interval $t_0$ is more than the time which the car needs to go through the critical section ($cs_1 - cs_0$), the car does not exist in the critical section while the signal is red.

Figure 2: A signal control system

Hybrid automata are models of hybrid systems with discrete and continuous behavior. We give a model of the above simple hybrid system according to the literature [8]. A hybrid automaton with polynomial dynamics is a tuple (*Loc*, *Lab*, *Edg*, *X*, *Init*, *Inv*, *Flow*, *Jump*) where *Loc* is a finite set of locations, *Lab* is a finite set of labels, $Edg \subseteq Loc \times Lab \times Loc$ is finite set of edges, *X* is a finite set of variables, *Init* and *Inv* give the initial condition

*Init*(*l*) and the invariant *Inv*(*l*) of location $l \in Loc$ as a polynomial constraints over $X$ respectively, *Flow* gives the flow constraint *Flow*(*l*) as a polynomial constraints over $X \cup X'$ where $X'$ is the set of the time derivative of variables $X$, and *Jump* gives the jump condition *Jump*(*e*) for edge *e* as a polynomial constraints over $X \cup X^+$ where $X^+$ refers to the updated values of the variables after the edge has been executed.

The semantics of a hybrid automaton is the transition system $(S, S_0, \rightarrow)$ where $S = \{(l, v) \in Loc \times \mathbb{R}^X \mid Inv(l) \text{ holds under } v\}$ is the state space, $S_0 = \{(l, v) \in Loc \times \mathbb{R}^X \mid Init(l) \text{ holds under } v\}$ is the initial space, and $\rightarrow$ is the transition relation defined as follows: $(l, v) \rightarrow_\sigma (k, w)$ if and only if (discrete transition) *Jump*(*e*) holds under $(v, w)$ for some $e = (l, \sigma, k) \in Edg$ where $w$ is the updated value by $e$, or (continuous transition) $l = k$ and $\sigma = tick_r$ for some $r \in \mathbb{R}^{\geq 0}$ such that *Flow*(*l*) holds under $(v, w)$ in the interval $[0, r]$ where $v$ and $w$ correspond to 0 and $r$.

Figure 3: A hybrid automaton of a signal control system

Figure 3 represents a hybrid automaton of our simple signal control system with a single car. Locations and edges are depicted as circles and arrows between them. Variables $X$ is given as $\{pos, now, l\}$. The initial values are all zero, that is, $Init(pos) = Init(now) = Init(l) = 0$, where *pos* and *now* stand for the position of the car and the current time, and $l$ is used for the signal interval control. $Loc = \{g, y, r\} \times \{0, 1\} \times \{0, 1\}$ is the set of locations where $(label, going, cs) \in Loc$ represents the location where the color of the signal is *label*, the car moves when *going* and the car exists in the critical section when *cs*. The top, the middle and the bottom labels in a location correspond to locations, flow conditions and invariants of the location respectively.

The invariants of locations are given by $STOP : pos' = 0 \land now' = 1$ or $GOING :$

$pos' = 1 \wedge now' = 1$, where $x'$ stands for the time derivative of $x$. Since $now' = 1$ holds in each location, the value of *now* always increases according to time advancing. Thus, *now* keeps the elapsed time from the initial state. The value of *pos* is unchanged when *STOP* holds, and increases when *GOING* holds. In the locations with the invariant $CS : cs_0 \leq pos \leq cs_1$, the car exists in the critical section. In the locations with $\overline{CS} : (0 \leq pos \leq cs_0) \vee cs_1 \leq pos$, the car does not exist in the critical section. Note that the borders $cs_0$ and $cs_1$ are included in both areas of $CS$ and $\overline{CS}$. The status can be changed at positions $cs_0$ and $cs_1$ only. Note that there are no locations satisfying that the signal color is not green, the car exists in the critical section, and the car stops.

The top-left set of locations $(g, \_, \_)$ represents those of the green signal. The bottom-right set $(y, \_, \_)$ and the bottom-left set $(r, \_, \_)$ represent those of the yellow and red signal. The labels of edges is given as $Lab = \{change, go, stop, in, out\}$. The edges $e$ between different color's locations are labeled by *change* and $Jump(e)$ is given as $CH : l \leq now \wedge l^+ = now + t_0$, which means that the edge can be executed when $l \leq now$ and then the value of $l$ is updated by $now + t_0$. We omit some of *change* and $CH$ in Figure 3. The labels *go* and *stop* change the behavior of the car: $(\_, 0, \_) \leftrightarrow (\_, 1, \_)$. The labels *in* and *out* change the status of the car about the critical section: $(\_, \_, 0) \leftrightarrow (\_, \_, 1))$. Note that the car does not enter the critical section when the signal is yellow or red. We omit unchanged variables in the jump condition, that is, $pos^+ = pos \wedge now^+ = now \wedge l^+ = l$ for the edges within same color's locations and $pos^+ = pos \wedge now^+ = now$ for the edges between different color's locations.

From the above conditions, there is no path $s_0 \to s_1 \to \cdots \to s_n$ such that $s_0 \in S_0$, $s_n = ((red, \_, \_), v)$ and $cs_0 < v(pos) < cs_1$, that is, the car does not exist in the critical section when the signal is red. We call it the safety property. In the following sections, we describe OTS/CafeOBJ specifications of the above hybrid automaton with a single car and with plural cars, and give a formal proof of the safety property.

### 3.2 An OTS/CafeOBJ specification of a signal control system with a single car

In our OTS model of the signal control system, there are three observations for discrete locations and three observations for continuous variables. Observations *color*, *going* and *cs* observes the value of elements of location (*label*, *going*, *cs*). Observations *now*, *loc* and *l* observes the value of those variables. There are two kinds of transitions in OTS models: discrete and continuous transitions. Discrete transitions *go*, *stop*, *in*, *out* and *change* correspond to edges. A continuous transition $tick_t$ advances the system's time by $t \in \mathbb{Q}^1$.

We give a system module `SIGNAL` which imports data modules `RAT` and `LABEL` given in the previous section. The following is a part of module `SIGNAL` specifying the initial state [2]:

```
bop now : Sys -> Rat {memo}
```

---

[1]We assume a temporal domain is dense, that is, there is a point between any pair of points. We use rational numbers for the clock values.

[2]Operation attribute `memo` is used only for effective computation with memorization in specification execution and does not have any effect on denotational semantics of specifications.

```
bop pos : Sys -> Rat {memo}
bop going : Sys -> Bool {memo}
bop cs : Sys -> Bool {memo}
bop color : Sys -> Label {memo}
bop l : Sys -> Rat {memo}

op init : -> Sys
eq now(init) = 0 .
eq pos(init) = 0 .
eq going(init) = false .
eq cs(init) = false .
eq color(init) = green .
eq l(init) = 0 .
```

The initial values of *now*, *pos* and *l* are defined as 0. The initial values of *going* and *cs* are false. The initial color is green. Module SIGNAL includes declaration ops cs0 cs1 t0 : -> Rat of constants $cs_0$, $cs_1$ and $t_0$.

Transition *change* is specified as follows:

```
bop c-change : Sys -> Bool {memo}
eq c-change(S) = l(S) <= now(S) .
ceq change(S) = S if not c-change(S) .

ceq color(change(S)) = next(color(S))
   if c-change(S) .
ceq l(change(S)) = now(S) + t0 if c-change(S) .

eq now(change(S)) = now(S) .
...
```

The effective condition c-change(S) is defined in the first equation such that $l \leq now$. The second (conditional) equation specifies that the state is unchanged if *change* is not effective. The third and fourth equations specifies the updated values of *color* and *l* such that *color* becomes the next color and *l* is set to $t_0$ time later. The remaining equations specify other variables are unchanged.

Transition *in* is specified as follows:

```
eq c-in(S) =
   (cs0 = pos(S) and color(S) = green) .
ceq in(S) = S if not c-in(S) .
ceq cs(in(S)) = true if c-in(S) .
eq now(in(S)) = now(S) .
...
```

The effective condition c-in(S) is defined in the first equation such that the car exists at $cs_0$ and the signal is green. When *in* is effective, cs(in(S)) becomes true, that is, location (\_, \_, 0) is changed into location (\_, \_, 1). The other discrete transitions are defined similarly.

Next, we specify the continuous transition. Time advancing $tick_r$ is described as follows:

```
bop now : Sys -> Rat {memo}
bop tick : Rat Sys -> Sys {memo}
var S : Sys
```

```
  var X : Rat
  ceq now(tick(X,S)) = now(S) + X if c-tick(X,S) .
  ceq pos(tick(X,S)) =
    (if going(S) then pos(S) + X else pos(S) fi)
      if c-tick(X,S) .
```

Term tick($r, s$) is the result state of applying $tick_r$ to state $s$. When $tick_r$ is done, the current time *now* increases by $r$. The position *pos* increases by $r$ if *going* is true, that is, the car moves forward.

The effective condition `c-tick` is given by invariants of the hybrid automaton.

```
eq c-tick(X,S) =
0 <= X and X <= cs1 - cs0
and
(cs(S) and going(S) implies
   cs0 <= pos(S) + X and pos(S) + X <= cs1)
and
(not cs(S) and going(S) implies
  pos(S) + X <= cs0 or cs1 <= pos(S) + X)
and
(not color(S) = green and cs(S) implies
 going(S))
and
(going(S) and cs0 < pos(S) + X and
 pos(S) + X <= cs1 implies cs(S))
and
(cs1 < pos(S) + X implies not cs(S)) .
```

Invariants in hybrid automata should always hold, which means that time cannot advance if the invariants do not hold. Thus, the effective condition `c-tick` is given as the conjunction of all invariants. For example, the invariant of location $(\_, 1, 1)$ is $CS$. The second and the third values of locations correspond to *going* and *cs*. Condition (`cs(S) and going(S) implies cs0 <= pos(S) + X and pos(S) + X <= cs1`) means that $cs_0 \leq pos^+$ and $pos^+ \leq cs_1$ hold whenever both *cs* and *going* are true. The effective condition of $tick_r$ should check invariant $CS$ by the updated value $pos^+$, which is $pos + r$. If future values violate invariants, time cannot advance.

### 3.3 An OTS/CafeOBJ specification of a signal control system with plural cars

Consider the case that more than one cars appear in our signal control system. The system is an example of multitask hybrid systems. In OTS models of multitask systems, observations and transitions related to processes are parameterized, for example, $pos_p$ and $go_p$ are an observation and a transition for a process $p$ respectively. In OTS/CafeOBJ specifications, multitask systems are described with the import of loose module PID. For example, $pos_p$ is given by `pos : Pid Sys -> Rat`.

We give a system module `MS` which imports data modules `RAT`, `LABEL` and `PID` for a signal control system with plural cars. Observations and initial state of module `MS` are specified as follows:

```
  bop now : Sys -> Rat {memo}
```

```
bop pos : Pid Sys -> Rat {memo}
bop going : Pid Sys -> Bool {memo}
bop cs : Pid Sys -> Bool {memo}
bop color : Sys -> Label {memo}
bop l : Sys -> Rat {memo}

op init : -> Sys
vars P : Pid
eq now(init) = 0 .
eq pos(P,init) = 0 .
eq going(P,init) = false .
eq cs(P,init) = false .
eq color(init) = green .
eq l(init) = 0 .
```

where observations `pos`, `going`, `cs` related to cars are parameterized. Equation `pos(P,init) = 0` means that the initial positions are zero for all cars P, for example.

The definition of discrete transitions are modified as follows:

```
vars P P' : Pid
eq c-in(P,S) =
  (cs0 = pos(P,S) and color(S) = green) .
ceq in(P,S) = S if not c-in(P,S) .
eq cs(P',in(P,S)) = P' = P or cs(P',S)  .
eq now(in(P,S)) = now(S) .
...
```

In the third equation above, term $\mathtt{cs}(p',\mathtt{in}(p,s))$ means that the value of $cs_{p'}$ at the result state of applying $in_p$ to state $s$ for processes $p$ and $p'$. Thus, when $p = p'$, it is true otherwise it is unchanged.

The effective condition of $tick_r$ should check all invariants for all processes. First, we parameterize `c-tick(P,X,S)` by processes P as follows:

```
eq c-tick(P,X,S) =
(cs(P,S) and going(P,S) implies
 cs0 <= pos(P,S) + X and pos(P,S) + X <= cs1)
and
(not cs(P,S) and going(P,S) implies
 pos(P,S) + X <= cs0 or cs1 <= pos(P,S) + X)
and
(not color(S) = green and cs(P,S) implies
 going(P,S))
and
(going(P,S) and cs0 < pos(P,S) + X and
 pos(P,S) + X <= cs1 implies cs(P,S))
and
(cs1 < pos(P,S) + X implies not cs(P,S)) .
```

If we fix the number of processes, it is easy to describe effective condition of $tick_r$. Consider the system with three processes `p1`, `p2` and `p3`. The effective condition `c-tick(X,S)` can be described as follows:

```
eq c-tick(X,S) =  0 <= X and X <= cs1 - cs0
and c-tick(p1,X,S) and c-tick(p2,X,S)
and c-tick(p3,X,S) .
```

We give a way to describe multitask hybrid systems in OTS/CafeOBJ specifications which denote hybrid automata without fixed numbers of processes. We extend a way to describe multitask real-time systems proposed in the literatures [5, 6].

We introduce a specification PSET of a set of processes:

```
mod* PSET{
  [Pid < PSet]
  op _ _ : PSet PSet -> PSet {assoc comm idem}
  op nil : -> PSet
  pred _in_ : Pid PSet
  vars P Q : Pid
  var PS : PSet
  eq (P in (P PS)) = true .
  eq (P in nil) = false .
  eq (P in Q) = (P = Q) .
  eq (P in (Q PS)) = (P = Q) or (P in PS) .
}
```

Sort PSet is declared as a super sort of Pid, and the sequence of two elements of PSet is also an element of PSet. Thus, a sequence of Pid is a term of PSet, for example, p1 p2 p3 is a term of PSet when p1, p2 and p3 are terms of Pid. Operation _in_ denotes the membership predicate on PSet. For example, p2 in p1 p2 p3 is true and p2 in p1 p3 is false.

We also introduce an observation $ps$ which is a set of active processes. A process becomes active when it moves. After a process is active, it is always active even if it stops. The initial value of $ps$ is empty. When a car $p$ starts to move, $ps$ is updated by $p\ ps$. The following is a part of description related to $ps$:

```
  bop ps : Sys -> PSet {memo}
  eq ps(init) = nil .
  eq ps(go(P,S)) = P ps(S) .
```

the effective condition c-tick is defined for $ps$. The effective condition of $tick_r$ is defined on $ps$ inductively as follows:

```
  op c-tick : PSet Rat Sys -> Bool {memo}

  eq c-tick(nil,X,S) = true .
  eq c-tick(P,X,S) = ...
  eq c-tick(P PS,X,S) = c-tick(P,X,S) and
                       c-tick(PS,X,S) .
  eq c-tick(X,S) =   0 <= X and X <= cs1 - cs0 and
                    c-tick(ps(S),X,S) .
```

By the first three equations, c-tick($ps, x, s$) is defined as the conjunction of c-tick($p, x, s$) for all processes $p$ in $ps$. For example, c-tick($ps, x, s$) is equivalent to c-tick(p1, $x, s$) and c-tick(p2, $x, s$) and c-tick(p3, $x, s$) when $ps$ = p1 p2 p3. The effective condition of $tick_r$ is defined by c-tick(ps(S),X,S).

## 4 Simulation of multitask hybrid systems

The following is an example of simulation by the reduction command for module MS:

```
open MS .
ops p1 p2 : -> Pid .
eq (p1 = p2) = false .
op st : Nat -> Sys .
eq cs0 = 5 .
eq cs1 = 10 .
eq t0 = 5 .
eq st(1) = tick(3,(go(p1,init))) .
eq st(2) = in(p1,tick(2,go(p2,st(1)))) .
eq st(3) = in(p2, tick(3,st(2))) .
eq st(4) = out(p1,tick(2,st(3))) .
eq st(5) = tick(3,st(4)) .
red pos(p1,st(1)) . red pos(p2,st(1)) .
...
red pos(p1,st(5)) . red pos(p2,st(5)) .
close
```

We give two processes $p_1$ and $p_2$ and concrete values for $cs_0 = 5$, $cs_1 = 10$ and $t_0 = 5$. State `st(1)` is the state after $p_1$ starts moving and three time unit passes from the initial state. State `st(2)` is the state after $p_2$ starts moving, two time unit passes and $p_1$ changes $cs$ into true from state `st(1)`. At state `st(2)`, the positions of $p_1$ and $p_2$ are five and two. In state `st(3)`, three time unit passes and $p_2$ changes $cs$ into true, where $(pos_{p_1}, pos_{p_2}) = (8, 5)$. In state `st(4)`, two time unit passes and $p_1$ changes $cs$ into false, where $(pos_{p_1}, pos_{p_2}) = (10, 7)$. In state `st(5)`, three time unit passes, where $(pos_{p_1}, pos_{p_2}) = (13, 10)$. CafeOBJ interpreter return expected values for the above codes.

Consider the following simulation:

```
eq st(11) = in(p1,tick(2,change(go(p2,st(1))))) .
eq st(12) = tick(3,(st(11))) .
eq st(13) = tick(3,stop(p1,st(11))) .
red now(st(12)) . red now(st(13)) .
```

From state `st(1)` with $(pos_{p_1}, pos_{p_2}) = (3, 0)$, $p_2$ starts moving, the signal is changed into yellow, two time unit passes and $p_1$ tries to change $cs$ into true in state `st(2)`. Since the signal is not green, $p_1$ cannot change $cs$ into true. State `st(11)` corresponds to location $(y, 1, 0)$ in Figure 3. State `st(12)` is obtained by applying $tick_3$ to `st(11)`, however time does not advance since the effective condition of $tick_3$ is false. To make the effective condition true, $p_1$ should stop. State `st(13)` is obtained by applying $stop_{p_1}$ and $tick_3$ to `st(11)`. which corresponds to location $(y, 0, 0)$. Then, time can advance. As we expected, `now(st(12))` is reduced into 5 and `now(st(13))` is reduced into 8.

Like above, we can test whether a described specification behaves as we expected by the reduction command for specific concrete values. Not only testing but also proving some property can be done by the reduction command.

## 5 Verification of multitask hybrid systems

In this section we give a formal proof of the safety property. We declare the relationship between constants $cs_0$, $cs_1$ and $t_0$ such that $0 < cs_0 < cs_1$ and $cs_1 - cs_0 \leq t_0$. Then, the specification denote all models satisfying the above condition including the test model in the previous section. First, we give a state predicate $inv_1(p, s)$ such that the car $p$ does not exist in the critical section when the signal is red at state $s$.

```
mod INV{
  pr(MS)
  pred inv1 : Pid Sys
  var P : Pid
  var S : Sys
  var X : Rat
  eq inv1(P,S) = not (color(S) = red and
      cs0 < pos(P,S) and pos(P,S) < cs1) .
}
```

If we prove $inv_1(s, p)$ for all processes $p$ and all reachable state $s$ from the initial state, the safety property holds. We denote the set of all reachable states by $RS$. We prove $\forall s \in RS. \forall p \in A_{\texttt{Pid}}. inv_1(p, s)$ by the induction on structure of reachable terms.

## 5.1 A proof passage for the induction basis

As the induction basis, the following proof passage is given to prove the initial state to satisfy $inv_1$.

```
open INV .
  op p : -> Pid .
  red inv1(p, init) .
close
```

CafeOBJ interpreter returns `true` for the above reduction command, which implies the induction basis holds, that is, $inv_1$ holds at the initial state.

## 5.2 Proof passages for the induction steps

In the induction step, we prove that each transition preserves the state predicate. We first assume an arbitrary state s satisfies inv1(p,s) for all processes p. Then, under the assumption we prove inv1(p,s') for the state s' obtained by any transition. The following is a template module for the induction step.

```
mod ISTEP{
  pr(INV)
  pred istep1 : Pid
  ops s s' : -> Sys
  var P : Pid
  eq istep1(P) = inv1(P,s) implies inv1(P,s') .
}
```

The proposition istep1(P) means that the proposition inv1(P,s') holds under the induction hypothesis inv1(P,s). The following is a part of the induction steps with respect to $tick_{t_1}$, where s' is obtained by applying $tick_{t_1}$ to s.

```
open ISTEP .
  op p : -> Pid .
  op t1 : -> Rat .
```

```
  eq s' = tick(t1,s) .
  red istep1(p) .
close
```

If the above reduction returns `true`, it guarantees that implication $inv_1(p, s) \Rightarrow inv_1(tick_{t_1}(p, s))$ holds. Unfortunately, it returns neither `true` nor `false` for the above proof passage. We need to give more information for proofs. A typical proof strategy is a case splitting by the effective condition as follows:

```
open ISTEP .
  op p : -> Pid .
  op t1 : -> Rat .
  eq c-tick(t1,s) = false .
  eq s' = tick(t1,s) .
  red istep1(p) .
close
```

```
open ISTEP .
  op p : -> Pid .
  op t1 : -> Rat .
  eq c-tick(t1,s) = true .
  eq s' = tick(t1,s) .
  red istep1(p) .
close
```

If the above two proof passages both return `true`, the original proof passage is satisfied, since $(c - tick(t_1, s) \Rightarrow istep1(p)) \land (\neg c - tick(t_1, s) \Rightarrow istep1(p)) \Rightarrow istep1(p)$. If it does not so, we proceed case-splitting more.

### 5.3 Lemma introduction

Repeating the process of case splitting, we may face `false` as a returned value of the reduction command. The following is such an example.

```
open ISTEP .
  op p : -> Pid .
  op t1 : -> Rat .
  ...
  eq cs(p,s) = true .
  ...
  eq cs0 <= pos(p,s) = false .
  eq s' = tick(t1,s) .
  red istep1(p) .
close
```

Since $cs_p$ is true, the position $pos_p$ is greater than or equal to $cs_0$. The equation `cs0 <= pos(p,s) = false` contradicts to it. Such a proof passage represents unreachable states.

In this case we introduce another appropriate safety property, called a lemma. We add the following lemma to the module `INV` and `ISTEP`:

```
  pred inv3 : Pid Sys
  eq inv3(P,S) = cs(P,S) implies
```

```
    cs0 <= pos(P,S) and pos(P,S) <= cs1 .

  pred istep3 : Pid
  eq istep3(P) = inv3(P,s) implies inv3(P,s') .
```

The lemma `inv3(P,S)` denotes that the position of car $p$ is between $cs_0$ and $cs_1$ whenever $cs_p$ is true. By replacing the reduction command by adding the lemma, we obtain `true` for the above proof passage.

```
red inv3(p,s) implies istep1(p) .
```

Proceeding the case splitting and introducing lemma, we obtain `true` for all the remaining proof passages for $inv_1$.

### 5.4 Proving lemma

We proved invariant $inv_1$ for all states reachable from the initial state by the induction scheme under the assumption of some lemmata. To complete the proof, we need to show (1) the lemmata hold for the initial state, and (2) the lemmata hold for result state of applying every transitions to states satisfying $inv_1$ and the lemmata. In the other words, we make a proof score of the conjunction of $inv_1 \wedge \cdots \wedge inv_n$, where the induction base is represented by $inv_1(p, init) \wedge \cdots \wedge inv_n(p, init)$ and the induction step is represented by $inv_1(p, s') \wedge \cdots \wedge inv_n(p, s')$ under the induction hypothesis $inv_1(p, s) \wedge \cdots \wedge inv_n(p, s)$. Note that in the previous section, we prove $inv_3(p, s) \Rightarrow (inv_1(p, s) \Rightarrow inv_1(p, s'))$. The formula is equivalent to $(inv_1(p, s) \wedge inv_3(p, s)) \Rightarrow inv_1(p, s')$. If we prove $(inv_1(p, s) \wedge inv_3(p, s)) \Rightarrow inv_3(p, s')$ for lemma $inv_3$, we obtain $(inv_1(p, s) \wedge inv_3(p, s)) \Rightarrow inv_1(p, s') \wedge inv_3(p, s')$.

To complete a proof of $inv_1$, we make seven lemmas and 136 proof passages[3], all of which return `true`. The following is the declaration of the lemmata `inv2`, `inv3`, `inv4`, `inv5`, `inv6` and `inv7`.

```
  eq inv1(P,S) = not (color(S) = red and
    cs0 < pos(P,S) and pos(P,S) < cs1) .
  eq inv2(P,S) = not (cs(P,S) and
    pos(P,S) < cs1 and color(S) = red) .
  eq inv3(P,S) = cs(P,S) implies
    cs0 <= pos(P,S) and pos(P,S) <= cs1 .
  eq inv4(P,S) = cs(P,S) and not color(S) = green
    and l(S) <= now(S) implies cs1 <= pos(P,S) .
  eq inv5(P,S) = cs(P,S) and not color(S) = green
    implies cs1 - pos(P,S) <= l(S) - now(S) .
  eq inv6(P,S) = cs(P,S) or cs0 = pos(P,S) or
    going(P,S) implies P in ps(S) .
  eq inv7(P,S) = cs0 < pos(P,S) and pos(P,S) < cs1
    implies cs(P,S) .
```

---

[3] Besides them, we need to add some lemmata which can be proved without induction on reachable states, e.g. `eq lemma1(P,X,S) = (P in ps(S)) and c-tick(X,S) implies c-tick(P,X,S)`. We proved it by the induction on structure of terms of sort `PSet`.

## 6  Related work

There are several tools for analyzing and verifying hybrid systems: HSolver[4], HyTech[5], KeYmaera[6], PHAVer[7] and so on. See the literature [8] for more details. One of the most relevant tools to our study is Maude a language and tool supporting specification description and verification based on rewriting logic [9]. Both Maude and CafeOBJ are algebraic specification languages and support user-defined abstract data type specifications, which is an advantage over the other tools for hybrid systems. Real-time Maude [10] is an extension of Maude which supports formal specification and analysis of real-time and hybrid systems. HI-Maude [11] is another extension of Maude which deals with a wider range of hybrid systems, called interacting hybrid systems. System modules in Maude are based on rewriting logic, where systems transitions are described by rewrite rules. Verification in Maude is based on exhaustive searching for reachable spaces obtained by the rewrite rules. In Maude, systems with discrete and continuous variables can be described, however, only discrete time domains obtained by time sampling strategies can be verified by search and model checking.

One of our advantages against these model checking approaches is that proof scores guarantee that verified properties hold for an arbitrary number of multiple processes. To make state spaces finite, model-checking approaches should restrict the size of the system to finite. Model checking is fully-automated. The proof score method is semi-automated and needs a human interaction to complete proofs. In the literature [12], automated support of making proof scores has been proposed. A case splitting phase may be automated, however, lemma discovery is heuristic and not easy to be automated.

## 7  Conclusion

We described an observational transition system of a simple signal control system with plural cars as an example of multitask hybrid systems, and verified some safety property by the proof score method. One of our future work is to apply the proposed method to practical applications of multitask hybrid systems, such as real-time operating systems, automotive control systems, intelligent transport systems, and so on.

## Acknowledgment

This work was supported by JSPS KAKENHI Grant Number JP19K11842.

## References

[1] R. Diaconescu and K. Futatsugi, CafeOBJ Report, World Scientific, Singapore, 1998.

---

[4] http://hsolver.sourceforge.net/
[5] https://ptolemy.berkeley.edu/projects/embedded/research/hytech/
[6] http://symbolaris.com/info/KeYmaera.html
[7] http://www-verimag.imag.fr/%7Efrehse/phaver_web/index.html

[2] K. Ogata, and K. Futatsugi, Proof scores in the OTS/CafeOBJ method, In: Najm, E., Nestmann, U., Stevens, P. (eds.) FMOODS 2003. LNCS, vol. 2884, pp. 170-184. Springer, Heidelberg,2003.

[3] K. Ogata and K. Futatsugi, Some tips on writing proof scores in the OTS/-CafeOBJ method, In: Futatsugi, K., Jouannaud, J.-P., Meseguer, J. (eds.) Goguen Festschrift. LNCS, vol. 4060, pp. 596-615. Springer, Heidelberg, 2006.

[4] K. Ogata and K. Futatsugi, Modeling and verification of real-time systems based on equations, Science of Computer Programming, 66(2), pp.162-180, Elsevier, 2007.

[5] S. Higashi, M. Nakamura, K. Sakakibara, C. Kojima and K. Ogata, Formal specification of multitask real-time systems using the OTS/CafeOBJ method, Proceedings of the SICE Annual Conference 2019 (SICE 2019), pp.122-125, 2019.

[6] M. Nakamura, S. Higashi, K. Sakakibara and K. Ogata, Formal verification of Fischer's real-time mutual exclusion protocol by the OTS/CafeOBJ method, Proceedings of the SICE Annual Conference 2020 (SICE 2020), pp.1210-1215, 2020

[7] Kazuhiro Ogata, Daigo Yamagishi, Takahiro Seino and Kokichi Futatsugi: Modeling and Verification of Hybrid Systems Based on Equations, Proceedings of the IFIP 18th World Computer Congress TC10 Working Conference on Distributed and Parallel Embedded Systems (DIPES 2004), Kluwer, pp.43-52, 2004.

[8] Laurent Doyen, Goran Frehse, George J. Pappas, and André Platzer, Verification of Hybrid Systems, Handbook of Model Checking, pp.1047-1110, Springer, 2018.

[9] P. C. Ölveczky and J. Meseguer, Semantics and Pragmatics of Real-Time Maude, Higher-Order and Symbolic Computation, Springer, 20, 1-2, pp.161-196, 2007.

[10] Daniela Lepri, Erika Ábrahám and Peter Csaba Ölveczky, Timed CTL Model Checking in Real-Time Maude, Rewriting Logic and Its Applications - 9th International Workshop, WRLA 2012, Lecture Notes in Computer Science 7571, pp.182-200, 2012

[11] Muhammad Fadlisyah and Peter Csaba Ölveczky, The HI-Maude Tool, Algebra and Coalgebra in Computer Science CALCO 2013, LNCS 8089, pp.322-327, Springer, 2013.

[12] D. Gaina, D. Lucanu, K. Ogata, K. Futatsugi, On Automation of OTS/CafeOBJ Method. Specification, Algebra, and Software, pp.578-602, 2014.

# A   Codes for the OTS/CafeOBJ specification of a signal control system with a single car

```
mod! LABEL{
  [Label]
  ops green red yellow : -> Label
  pred _=_ : Label Label {comm}
  op next : Label -> Label
  var L : Label
  eq (L = L) = true .
  eq (green = red) = false .
  eq (green = yellow) = false .
  eq (red = yellow) = false .

  eq next(red) = green .
  eq next(green) = yellow .
  eq next(yellow) = red .
}

mod* SIGNAL{
  pr(RAT + LABEL)
  *[Sys]*
  ops cs0 cs1 t0 : -> Rat
  op init : -> Sys

  bop now : Sys -> Rat {memo}
  bop pos : Sys -> Rat {memo}

  bop going : Sys -> Bool {memo}
  bop cs : Sys -> Bool {memo}
  bop color : Sys -> Label {memo}
  bop l : Sys -> Rat {memo}

  bop tick : Rat Sys -> Sys {memo}
  bop go : Sys -> Sys {memo}
  bop stop : Sys -> Sys {memo}
  bop in : Sys -> Sys {memo}
  bop out : Sys -> Sys {memo}
  bop change : Sys -> Sys {memo}

  op c-tick : Rat Sys -> Bool {memo}
  bop c-change : Sys -> Bool {memo}
  bop c-in : Sys -> Bool {memo}
  bop c-out : Sys -> Bool {memo}

  var S : Sys
  vars X Y : Rat
```

```
eq now(init) = 0 .
eq pos(init) = 0 .
eq going(init) = false .
eq cs(init) = false .
eq color(init) = green .
eq l(init) = 0 .

eq c-tick(X,S) =
0 <= X and X <= cs1 - cs0 and
(cs(S) and going(S) implies cs0 <= pos(S) + X and pos(S) + X <= cs1) and
(not cs(S) and going(S) implies pos(S) + X <= cs0 or cs1 <= pos(S) + X) and
(not color(S) = green and cs(S) implies going(S)) and
(going(S) and cs0 < pos(S) + X and pos(S) + X <= cs1 implies cs(S)) and
(cs1 < pos(S) + X implies not cs(S)) .

ceq tick(X,S) = S if not c-tick(X,S) .
ceq now(tick(X,S)) = now(S) + X if c-tick(X,S) .
ceq pos(tick(X,S)) = (if going(S) then pos(S) + X else pos(S) fi) if c-tick(X,S) .
eq going(tick(X,S)) = going(S) .
eq cs(tick(X,S)) = cs(S) .
eq color(tick(X,S)) = color(S) .
eq l(tick(X,S)) = l(S) .

eq now(go(S)) = now(S) .
eq pos(go(S)) = pos(S) .
eq going(go(S)) = true .  -- (*)
eq cs(go(S)) = cs(S) .
eq color(go(S)) = color(S) .
eq l(go(S)) = l(S) .

eq now(stop(S)) = now(S) .
eq pos(stop(S)) = pos(S) .
eq going(stop(S)) = false .   -- (*)
eq cs(stop(S)) = cs(S) .
eq color(stop(S)) = color(S) .
eq l(stop(S)) = l(S) .

eq c-in(S) = (cs0 = pos(S) and color(S) = green) .
ceq in(S) = S if not c-in(S) . -- (*)
eq now(in(S)) = now(S) .
eq pos(in(S)) = pos(S) .
eq going(in(S)) = going(S) .
ceq cs(in(S)) = true if c-in(S) .        -- (*)
eq color(in(S)) = color(S) .
```

```
  eq l(in(S)) = l(S) .

  eq c-out(S) = (pos(S) = cs1) .
  ceq out(S) = S if not c-out(S) . -- (*)
  eq now(out(S)) = now(S) .
  eq pos(out(S)) = pos(S) .
  eq going(out(S)) = going(S) .
  ceq cs(out(S)) = false if not c-out(S) . -- (*)
  eq color(out(S)) = color(S) .
  eq l(out(S)) = l(S) .

  eq c-change(S) = l(S) <= now(S) .
  ceq change(S) = S if not c-change(S) .
  eq now(change(S)) = now(S) .
  eq pos(change(S)) = pos(S) .
  eq going(change(S)) = going(S) .
  ceq color(change(S)) = next(color(S)) if c-change(S) .
  eq cs(change(S)) = cs(S) .
  ceq l(change(S)) = now(S) + t0 if c-change(S) .
}
```

## B  Codes for the OTS/CafeOBJ specification of a signal control system with plural cars

```
mod* PID{
  [Pid]
  pred _=_ : Pid Pid {comm}
  vars P Q : Pid
  eq (P = P) = true .
}
mod* PSET{
  [Pid < PSet]
  op _ _ : PSet PSet -> PSet {assoc comm idem}
  op nil : -> PSet
  pred _in_ : Pid PSet
  vars P Q : Pid
  var PS : PSet
  eq (P in (P PS)) = true .
  eq (P in nil) = false .
  eq (P in Q) = (P = Q) .
  eq (P in (Q PS)) = (P = Q) or (P in PS) .
```

}

```
mod* MS{
  pr(PSET + RAT + LABEL)
  *[Sys]*
  bop now : Sys -> Rat {memo}
  bop pos : Pid Sys -> Rat {memo}

  bop going : Pid Sys -> Bool {memo}
  bop cs : Pid Sys -> Bool {memo}
  bop color : Sys -> Label {memo}
  bop l : Sys -> Rat {memo}
  bop ps : Sys -> PSet {memo}

  bop tick : Rat Sys -> Sys {memo}
  bop go : Pid Sys -> Sys {memo}
  bop stop : Pid Sys -> Sys {memo}
  bop in : Pid Sys -> Sys {memo}
  bop out : Pid Sys -> Sys {memo}
  bop change : Sys -> Sys {memo}

  op c-tick : Rat Sys -> Bool {memo}
  bop c-change : Sys -> Bool {memo}
  bop c-in : Pid Sys -> Bool {memo}
  bop c-out : Pid Sys -> Bool {memo}

  op init : -> Sys
  ops cs0 cs1 t0 : -> Rat
  var S : Sys
  vars X Y Z : Rat
  vars P P' : Pid
  vars PS : PSet

  eq now(init) = 0 .
  eq pos(P,init) = 0 .
  eq going(P,init) = false .
  eq cs(P,init) = false .
  eq color(init) = green .
  eq l(init) = 0 .
  eq ps(init) = nil .

  op c-tick : PSet Rat Sys -> Bool {memo}
```

```
eq c-tick(nil,X,S) = true .
eq c-tick(P,X,S) =
(cs(P,S) and going(P,S) implies cs0 <= pos(P,S) + X and pos(P,S) + X <= cs1) and
(not cs(P,S) and going(P,S) implies pos(P,S) + X <= cs0 or cs1 <= pos(P,S) + X) and
(not color(S) = green and cs(P,S) implies going(P,S)) and
(going(P,S) and cs0 < pos(P,S) + X and pos(P,S) + X <= cs1 implies cs(P,S)) and
(going(P,S) and cs1 < pos(P,S) + X implies not cs(P,S)) .

eq c-tick(P PS,X,S) = c-tick(P,X,S) and c-tick(PS,X,S) .

eq c-tick(X,S) =  0 <= X and X <= cs1 - cs0 and c-tick(ps(S),X,S) .

ceq tick(X,S) = S if not c-tick(X,S) .
ceq now(tick(X,S)) = now(S) + X if c-tick(X,S) .
ceq pos(P,tick(X,S)) = (if going(P,S) then pos(P,S) + X else pos(P,S) fi)
if c-tick(X,S) .
eq going(P,tick(X,S)) = going(P,S) .
eq cs(P,tick(X,S)) = cs(P,S) .
eq color(tick(X,S)) = color(S) .
eq l(tick(X,S)) = l(S) .
eq ps(tick(X,S)) = ps(S) .

eq now(go(P,S)) = now(S) .
eq pos(P',go(P,S)) = pos(P',S) .
eq going(P',go(P,S)) = P' = P or going(P',S) . -- (*)
eq cs(P',go(P,S)) = cs(P',S) .
eq color(go(P,S)) = color(S) .
eq l(go(P,S)) = l(S) .
eq ps(go(P,S)) = P ps(S) .

eq now(stop(P,S)) = now(S) .
eq pos(P',stop(P,S)) = pos(P',S) .
eq going(P',stop(P,S)) = (not P' = P) and going(P',S) . -- (*)
eq cs(P',stop(P,S)) = cs(P',S) .
eq color(stop(P,S)) = color(S) .
eq l(stop(P,S)) = l(S) .
eq ps(stop(P,S)) = ps(S) .

eq c-in(P,S) = (cs0 = pos(P,S) and color(S) = green) .
ceq in(P,S) = S if not c-in(P,S) . -- (*)
eq now(in(P,S)) = now(S) .
eq pos(P',in(P,S)) = pos(P',S) .
eq going(P',in(P,S)) = going(P',S) .
eq cs(P',in(P,S)) = P' = P or cs(P',S) .          -- (*)
eq color(in(P,S)) = color(S) .
eq l(in(P,S)) = l(S) .
```

```
  eq ps(in(P,S)) = ps(S) .

  eq c-out(P,S) = (pos(P,S) = cs1) .
  ceq out(P,S) = S if not c-out(P,S) . -- (*)
  eq now(out(P,S)) = now(S) .
  eq pos(P',out(P,S)) = pos(P',S) .
  eq going(P',out(P,S)) = going(P',S) .
  eq cs(P',out(P,S)) = (not P' = P) and cs(P',S) . -- (*)
  eq color(out(P,S)) = color(S) .
  eq l(out(P,S)) = l(S) .
  eq ps(out(P,S)) = ps(S) .

  eq c-change(S) = l(S) <= now(S) .
  ceq change(S) = S if not c-change(S) .
  eq now(change(S)) = now(S) .
  eq pos(P',change(S)) = pos(P',S) .
  eq going(P',change(S)) = going(P',S) .
  eq cs(P',change(S)) = cs(P',S) .
  ceq color(change(S)) = next(color(S)) if c-change(S) .
  ceq l(change(S)) = now(S) + t0 if c-change(S) .
  eq ps(change(S)) = ps(S) .

-- constants
  eq 0 < cs0 = true .
  eq 0 <= cs0 = true .
  eq cs0 < cs1 = true .
  eq cs1 <= 0 = false .
  eq cs0 <= cs1 = true .
  eq cs1 <= cs0 = false .
  eq 0 < t0 = true .
  eq cs1 + - cs0 <= t0 = true .
-- Lemma <=
  eq (X:Rat < Y:Rat) = not (Y <= X) .
  eq (X:Rat <= X) = true .
  ceq (X + Y <= X) = false if 0 < Y .
  ceq (X <= Y + Z) = true if 0 <= Y and X <= Z .
  eq - (X + Y) = - X + - Y .
  eq - X + X = 0 .
  ceq (X + Y <= X + Z) = true if Y <= Z .
}
```

## C  Codes for verification

```
mod INV{
```

```
    pr(MS)
    pred inv1 : Pid Sys
    pred inv2 : Pid Sys
    pred inv3 : Pid Sys
    pred inv4 : Pid Sys
    pred inv5 : Pid Sys
    pred inv6 : Pid Sys
    pred inv7 : Pid Sys
    pred lemma1 : Pid Rat Sys

    var P : Pid
    var S : Sys
    var X : Rat
    eq inv1(P,S) = not (color(S) = red and cs0 < pos(P,S) and pos(P,S) < cs1) .
    eq inv2(P,S) = not (cs(P,S) and pos(P,S) < cs1 and color(S) = red) .
    eq inv3(P,S) = cs(P,S) implies cs0 <= pos(P,S) and pos(P,S) <= cs1 .
    eq inv4(P,S) = cs(P,S) and not color(S) = green and l(S) <= now(S) implies
                   cs1 <= pos(P,S) .
    eq inv5(P,S) = cs(P,S) and not color(S) = green implies
                   cs1 - pos(P,S) <= l(S) - now(S) .
    eq inv6(P,S) = cs(P,S) or cs0 = pos(P,S) or going(P,S) implies P in ps(S) .
    eq inv7(P,S) = cs0 < pos(P,S) and pos(P,S) < cs1 implies cs(P,S) .

    eq lemma1(P,X,S) = (P in ps(S)) and c-tick(X,S) implies c-tick(P,X,S) .
}

mod ISTEP{
  pr(INV)
  pred istep1 : Pid
  pred istep2 : Pid
  pred istep3 : Pid
  pred istep4 : Pid
  pred istep5 : Pid
  pred istep6 : Pid
  pred istep7 : Pid

  ops s s' : -> Sys
  var P : Pid
  var X : Rat
  eq istep1(P) = inv1(P,s) implies inv1(P,s') .
  eq istep2(P) = inv2(P,s) implies inv2(P,s') .
  eq istep3(P) = inv3(P,s) implies inv3(P,s') .
  eq istep4(P) = inv4(P,s) implies inv4(P,s') .
  eq istep5(P) = inv5(P,s) implies inv5(P,s') .
  eq istep6(P) = inv6(P,s) implies inv6(P,s') .
  eq istep7(P) = inv7(P,s) implies inv7(P,s') .
```

}